\documentclass[a4paper]{jpconf}
\usepackage{graphicx}
\newcommand{\bpartial}{\mbox{\boldmath $\partial$}}
\newcommand{\balpha}{\mbox{\boldmath $\alpha$}}
\newcommand{\bnabla}{\mbox{\boldmath $\nabla$}}
\newcommand{\bsigma}{\mbox{\boldmath $\sigma$}}

\begin{document}
\title{Self-adjointness and the Casimir effect with confined 
quantized spinor matter}

\author{Yurii A Sitenko$^{1,2}$}

\address{$^1$ Bogolyubov Institute for Theoretical Physics, National Academy of Sciences of Ukraine, \\
14-b Metrologichna Str., 03680 Kyiv, Ukraine}

\address{$^2$ Institute for Theoretical Physics, University of Bern,
Sidlerstrasse 5, CH-3012 Bern, Switzerland}

\ead{yusitenko@bitp.kiev.ua}

\begin{abstract}
A generalization of the MIT bag boundary condition for spinor matter is proposed basing on the 
requirement that the Dirac hamiltonian operator be self-adjoint. An influence of a background 
magnetic field on the vacuum of charged spinor matter confined between two parallel 
material plates is studied. Employing the most general set of boundary conditions at the plates 
in the case of the uniform magnetic field directed orthogonally to the plates, we find the pressure 
from the vacuum onto the plates. In physically plausible situations, the Casimir effect is shown to 
be repulsive, independently of a choice of boundary conditions and of a distance between the
plates.
\end{abstract}

\section{Introduction}

The self-adjointness of operators of physical observables in quantum mechanics is required by general principles of 
comprehensibility and mathematical consistency, see, e.g., \cite{Akhi,Nai}. To put it 
simply, a multiple action is well defined for a self-adjoint operator only, allowing for the construction of functions of 
the operator, such as resolvent, evolution, heat kernel and zeta-function operators, with further implications upon second
quantization. 

The mathematical demand for the self-adjointness of a basic operator acting on functions defined in a bounded spatial region 
is a somewhat more general than the physical demand for the confinement of quantized matter fields inside this region. The 
concept of confined matter fields is quite familiar in the context of condensed matter physics: collective excitations 
(e.g., spin waves and phonons) exist only inside material objects and do not spread outside. In the context 
of quantum electrodynamics, if one is interested in the effect of a classical background magnetic field on the vacuum of 
the quantized electron-positron matter, then the latter should be considered as confined to the spatial region between 
the sources of the magnetic field, as long as collective quasielectronic excitations inside a magnetized material differ from 
electronic excitations in the vacuum. It should be noted in this respect that the effect of the background electromagnetic 
field on the vacuum of quantized charged matter was studied for eight decades almost, see \cite{Hei2,Schw} and a review in 
\cite{Dit}. However, the concern was for the case of a background field filling the whole (infinite) space, that 
is hard to be regarded as realistic. The case of both the background and quantized fields confined to a bounded spatial 
region with boundaries serving as sources of the background field looks much more physically plausible, it can even be 
regarded as realizable in laboratory. Moreover, there is no way to detect the energy density that is induced in the vacuum 
in the first case, whereas the pressure from the vacuum onto the boundaries, resulting in the second case, is in principle 
detectable.

In view of the above, an issue of a choice of boundary conditions ensuring the confinement of the quantized matter gains a crucial significance. It seems that a quest for such boundary conditions was initiated in 
the context of a model description of hadrons as composites of quarks and gluons \cite{Cho1,Cho2}. If an hadron 
is an extended object occupying spatial region $\Omega$ bounded by surface $\partial{\Omega}$, then the condition that the quark 
matter field be confined inside the hadron is formulated as
$$
\mathbf{n}\cdot\mathbf{J}(\mathbf{r})|_{\mathbf{r}
\in
\partial{\Omega}}=0,\eqno(1)
$$
where $\mathbf{n}$ is the unit normal to the boundary surface, and 
$\mathbf{J}(\mathbf{r})={\psi}^{\dag}(\mathbf{r})\balpha\psi(\mathbf{r})$, $\mathbf{r}
\in \Omega$, with $\psi(\mathbf{r})$ being the quark matter field (an appropriate condition is also formulated for the gluon 
matter field). Condition (1) should be resolved to take the form of a boundary condition that is linear in $\psi(\mathbf{r})$, and an immediate 
way of such a resolution is known as the MIT bag boundary condition \cite{Joh},
$$
[I + {\rm i}\beta(\mathbf{n}\cdot\balpha)]\psi(\mathbf{r})|_{\mathbf{r}\in
\partial{\Omega}}=0 \eqno(2)
$$
(${\alpha}^1$, ${\alpha}^2$, ${\alpha}^3$ and ${\beta}$ are the generating elements of the Dirac-Clifford algebra), but it is 
needless to say that this way is not a unique one. The most general boundary condition that is linear in 
$\psi(\mathbf{r})$ in the case of a simply-connected boundary involves four arbitrary parameters \cite{Wie}, and the explicit 
form of this boundary condition has been given \cite{Si1}; the 
condition is compatible with the self-adjointness of the Dirac hamiltonian operator, and its four parameters can 
be interpreted as the self-adjoint extension parameters. In the present work, I follow the lines of works \cite{Wie, Si1} 
by proposing a different, embracing more cases, form of the four-parameter generalization of the MIT bag boundary condition.

Thus, let us consider in general the quantized spinor matter field that is confined to the
three-dimensional spatial region $\Omega$ bounded by the two-dimensional surface $\partial{\Omega}$. To study 
a response of the vacuum to the background magnetic field, we restrict ourselves to the case of the boundary 
consisting of two parallel planes; the magnetic field is assumed to be uniform and orthogonal to the planes. 
Such a spatial geometry is typical for the remarkable macroscopic quantum phenomenon which yields the attraction 
(negative pressure) between two neutral plates and which is known 
as the Casimir effect \cite{Cas1}, see review in \cite{Bor}. The conventional Casimir effect is due to vacuum 
fluctuations of the quantized electromagnetic field obeying certain boundary conditions at the bounding plates, and 
a choice of boundary conditions is physically motivated by material properties of the plates (for instance, metallic 
or dielectric, see, e.g., \cite{Bor}). Such a motivation is lacking for the case of vacuum fluctuations of the 
quantized spinor matter field. That is why there is a necessity in the last case to start from the most general set of 
mathematically acceptable (i.e. compatible with the self-adjointness) boundary conditions. Further follow physical constraints 
that the spinor matter be confined within the plates and that the spectrum of the wave number vector in the direction which is 
orthogonal to the plates be unambiguously (although implicitly) determined. Employing these mathematical and physical 
restrictions, I consider the generalized Casimir effect which is due to vacuum fluctuations of the quantized spinor matter 
field in the presence of the background magnetic field; the pressure from the vacuum onto the bounding plates will be found.

\section{Self-adjointness and boundary conditions}

Defining a scalar product as $(\tilde{\chi},\chi)=\int\limits_{\Omega}{\rm
d}^3r\,\tilde{\chi}^{\dag}\chi$, we get, using integration by parts,
$$
(\tilde{\chi},H\chi)=(H^{\dag}\tilde{\chi},\chi)- {\rm
i}\int\limits_{\partial{\Omega}}{\rm
d}\mathbf{s}\cdot\tilde{\chi}^{\dag}\balpha\chi,\eqno(3)
$$
where
$$
H=H^{\dag}=-{\rm
i}\balpha\cdot\bnabla+\beta{m} \eqno(4)
$$
is the formal expression for the Dirac hamiltonian operator and ${\bnabla}$ is the covariant derivative involving both
the affine and bundle connections (natural units ${\hbar}=c=1$ are used). Operator $H$ is
Hermitian (or symmetric in mathematical parlance),
$$
(\tilde{\chi},H\chi)=(H^{\dag}\tilde{\chi},\chi),\eqno(5)
$$
if
$$
\int\limits_{\partial{\Omega}}{\rm
d}\mathbf{s}\cdot\tilde{\chi}^{\dag}\balpha\chi
= 0. \eqno(6)
$$
The latter condition can be satisfied in various ways by imposing
different boundary conditions for $\chi$ and $\tilde{\chi}$.
However, among the whole variety, there may exist a possibility
that a boundary condition for $\tilde{\chi}$ is the same as that
for $\chi$; then the domain of definition of $H^{\dag}$ (set of
functions $\tilde{\chi}$) coincides with that of $H$ (set of
functions $\chi$), and operator $H$ is self-adjoint. The
action of a self-adjoint operator results in functions belonging
to its domain of definition only, and, therefore, a multiple
action and functions of such an operator can be consistently defined.

Condition (6) is certainly fulfilled when the integrand in (6)
vanishes, i.e.
$$
\tilde{\chi}^{\dag}(\mathbf{n}\cdot\balpha)\chi|_{\mathbf{r}
\in \partial{\Omega}}=0. \eqno(7)
$$
To fulfill the latter condition, we impose the same
boundary condition for $\chi$ and $\tilde{\chi}$ in the form
$$
\chi|_{\mathbf{r}\in \partial{\Omega}}=K\chi|_{\mathbf{r}\in
\partial{\Omega}},\quad \tilde{\chi}|_{\mathbf{r}\in
\partial{\Omega}}=K\tilde{\chi}|_{\mathbf{r}\in \partial{\Omega}},\eqno(8)
$$
where $K$ is a matrix (element of the Dirac-Clifford algebra) which is
determined by two conditions:
$$
K^{2}=I \eqno(9)
$$
and
$$
K^{\dag}(\mathbf{n}\cdot\balpha)K=-\mathbf{n}\cdot\balpha. \eqno(10)
$$
It should be noted that, in addition to (7), the following
combination of $\chi$ and $\tilde{\chi}$ is also vanishing at the
boundary:
$$
\tilde{\chi}^{\dag}(\mathbf{n}\cdot\balpha)K\chi|_{\mathbf{r}
\in
\partial{\Omega}}=\tilde{\chi}^{\dag}K^{\dag}(\mathbf{n}\cdot\balpha)\chi|_{\mathbf{r}
\in \partial{\Omega}}=0. \eqno(11)
$$
Using the standard representation for the Dirac matrices,
$$
\beta=\left(\begin{array}{cc}
I&0\\
0&-I
\end{array}\right),\qquad
\balpha=\left(\begin{array}{cc}
0&\bsigma\\
\bsigma&0
\end{array}\right) \eqno(12)
$$
($\sigma^{1},\sigma^{2}$ and $\sigma^{3}$ are the Pauli matrices),
one can get
$$
K=\left(\begin{array}{cc}
0&{\varrho}^{-1}\\
\varrho&0
\end{array}\right), \eqno(13)
$$
where condition
$$
(\mathbf{n}\cdot\bsigma)\varrho=-{\varrho}^{\dag}(\mathbf{n}\cdot\bsigma)
\eqno(14)
$$
defines $\varrho$ as a rank-2 matrix depending on four arbitrary
parameters  \cite{Wie}. An explicit form for matrix $K$ is \cite{Si1}
$$
K=\frac{(1+u^2-v^2-{\mathbf{t}}^2)\beta+(1-u^2+v^2+{\mathbf{t}}^2)I}{2{\rm
i}(u^2-v^2-{\mathbf{t}}^2)}(u\mathbf{n}\cdot\balpha+v\beta\gamma^{5}-{\rm
i}\mathbf{t}\cdot\balpha),\eqno(15)
$$
where $\gamma^{5}={\rm i}\alpha^1\alpha^2\alpha^3$, and
$\mathbf{t}=(t^1,t^2)$ is a two-dimensional vector which is
tangential to the boundary, $\mathbf{t}\cdot\mathbf{n}=0$. Matrix $K$ is Hermitian in 
two cases only when it takes forms
$$
K_{+}=-{\rm i}\beta(\mathbf{n}\cdot\balpha) \quad
(u=1,\quad v=0, \quad \mathbf{t}=0) \eqno(16)
$$
and
$$
K_{-}={\rm i}v\beta\gamma^{5}+\mathbf{t}\cdot\balpha 
\quad (u=0,\quad v^2+{\mathbf{t}}^2=1). \eqno(17)
$$
Matrix $K_{+}$ (16) corresponds to the choice of the standard MIT bag boundary
condition \cite{Joh}, cf. (2),
$$
(I-K_{+})\chi|_{\mathbf{r}\in
\partial{\Omega}}=(I-K_{+})\tilde{\chi}|_{\mathbf{r}\in
\partial{\Omega}}=0, \eqno(18)
$$
when relation (11) takes form
$$
\tilde{\chi}^{\dag}\beta\chi|_{\mathbf{r}\in
\partial{\Omega}}=0.\eqno(19)
$$

It is instructive to go over from off-diagonal matrix $K$ (15) to Hermitian matrix $\tilde{K}$,  
presenting boundary condition (8) as 
$$
\chi|_{\mathbf{r}\in \partial{\Omega}}=\tilde{K}\chi|_{\mathbf{r}\in
\partial{\Omega}},\quad \tilde{\chi}|_{\mathbf{r}\in
\partial{\Omega}}=\tilde{K}\tilde{\chi}|_{\mathbf{r}\in \partial{\Omega}},\eqno(20)
$$
with $\tilde{K}=\tilde{K}^{\dag}$ determined by conditions
$$
\tilde{K}^{2}=I \eqno(21)
$$
and
$$
[\tilde{K},\mathbf{n}\cdot\balpha]_{+}=0.\eqno(22)
$$
This transition 
is implemented with the use of the block-diagonal Hermitian matrix, 
$$
N=\left(\begin{array}{cc}
               \nu_1 & 0 \\
               0 & \nu_2
             \end{array}\right),
\qquad  \nu_1^{\dag}=\nu_1,\qquad  \nu_2^{\dag}=\nu_2,\eqno(23)
$$
obeying condition
$$
(I-N)K=K^{\dag}(I-N); \eqno(24)
$$
namely, the result is
$$
\tilde{K}=(I-N)K+N. \eqno(25)
$$
Using parametrization
$$
u=-\frac{\sin\tilde{\varphi}}{\cos\varphi\cos\theta+\cos\tilde{\varphi}}, \quad
v=\frac{\sin\varphi\cos\theta}{\cos\varphi\cos\theta+\cos\tilde{\varphi}},
$$
$$
t^1=\frac{\sin\theta\cos\eta}{\cos\varphi\cos\theta+\cos\tilde{\varphi}}, \quad
t^2=\frac{\sin\theta\sin\eta}{\cos\varphi\cos\theta+\cos\tilde{\varphi}},
$$
$$
-\pi/2 < \varphi \leq \pi/2, \quad -\pi/2 \leq \tilde{\varphi} < \pi/2,
\quad 0\leq\theta<\pi, \quad 0\leq\eta<2\pi, \eqno(26)
$$
one gets
$$
K={\rm i}\frac{\beta\cos\varphi\cos\theta+I\cos\tilde{\varphi}}{\cos^2\varphi\cos^2\theta-\cos^2\tilde{\varphi}}
[\mathbf{n}\cdot\balpha\sin\tilde{\varphi}-\beta\gamma^{5}\sin\varphi\cos\theta+{\rm
i}(\alpha^{1}\cos\eta+\alpha^{2}\sin\eta)\sin\theta)],\eqno(27)
$$
where
$$
[\mathbf{n}\cdot\balpha,\,\alpha^1]_{+}=
[\mathbf{n}\cdot\balpha,\,\alpha^2]_{+}=[\alpha^1,\alpha^2]_{+}=0.
\eqno(28)
$$
Then matrix $N$ takes form
$$
N=\beta\cos\varphi\cos\tilde{\varphi}\cos\theta
-\beta\gamma^{5}(\mathbf{n}\cdot\balpha)\sin\varphi\sin\tilde{\varphi}\cos\theta
+{\rm i}(\alpha^{1}\cos\eta+\alpha^{2}\sin\eta)(\mathbf{n}\cdot\balpha)\sin\tilde{\varphi}\sin\theta, \eqno(29)
$$
and one gets
$$
\tilde{K}=[\beta{\rm
e}^{{\rm
i}\varphi\gamma^{5}}\cos\theta+(\alpha^{1}\cos\eta+\alpha^{2}\sin\eta)\sin\theta]{\rm
e}^{{\rm
i}\tilde{\varphi}\mathbf{n}\cdot\balpha}; \eqno(30)
$$
in particular,
$$
K_{+}=\tilde{K}|_{\varphi=0, \, \tilde{\varphi}=-\pi/2, \, \theta=0}, \quad
K_{-}=\tilde{K}|_{\varphi=\pi/2, \, \tilde{\varphi}=0}.\eqno(31)
$$

Thus, the explicit form of the boundary condition ensuring the
self-adjointness of operator $H$ (4) is
$$
\left\{I - [\beta{\rm
e}^{{\rm
i}\varphi\gamma^{5}}\cos\theta+(\alpha^{1}\cos\eta+\alpha^{2}\sin\eta)\sin\theta]{\rm
e}^{{\rm
i}\tilde{\varphi}\mathbf{n}\cdot\balpha}\right\}\chi|_{\mathbf{r}\in
\partial{\Omega}}=0 \eqno(32)
$$
(the same condition is for $\tilde{\chi}$), and relation (11) takes
form
$$
\tilde{\chi}^{\dag}[\beta{\rm
e}^{{\rm
i}\varphi\gamma^{5}}\cos\theta+(\alpha^{1}\cos\eta+\alpha^{2}\sin\eta)\sin\theta]{\rm
e}^{{\rm
i}(\tilde{\varphi} + \pi/2)\mathbf{n}\cdot\balpha}\chi|_{\mathbf{r}\in
\partial{\Omega}}=0.\eqno(33)
$$
Four parameters of boundary condition (32), $\varphi, \tilde{\varphi}, \theta$ and $\eta$, 
can be interpreted as the self-adjoint extension parameters. It should be emphasized that the values 
of these parameters vary in general from point to point of the
boundary. In this respect the ``number'' of self-adjoint extension
parameters is in fact infinite, moreover, it is not countable but
is of power of a continuum. This distinguishes the case of an
extended boundary from the case of an excluded point (contact
interaction), when the number of self-adjoint extension parameters
is finite, being equal to $n^2$ for the deficiency index equal to
\{$n,n$\} (see, e.g., \cite{Akhi}).

In the context of the Casimir effect, one usually considers
spatial region $\Omega$ with a disconnected boundary consisting of
two connected components, $\partial{\Omega} =
\partial{\Omega}^{(+)}\bigcup\partial{\Omega}^{(-)}$. Choosing
coordinates $\mathbf{r}=(x,y,z)$ in such a way that $x$ and $y$
are tangential to the boundary, while $z$ is normal to it, we
identify the position of $\partial{\Omega}^{(\pm)}$ with, say,
$z=\pm{a/2}$. In general, there are 8 self-adjoint extension parameters: $\varphi_+$, $\tilde{\varphi}_+$, $\theta_+$  and
$\eta_+$ corresponding to $\partial{\Omega}^{(+)}$ and $\varphi_-$, $\tilde{\varphi}_-$, $\theta_-$  and $\eta_-$ corresponding to
$\partial{\Omega}^{(-)}$. However, if some symmetry is present,
then the number of self-adjoint extension parameters is diminished. For instance, if the boundary consists of two parallel planes, then the
cases differing by the values of $\eta_+$ or $\eta_-$ are physically indistinguishable, since they are related by a rotation around a normal
to the boundary.  To avoid this unphysical degeneracy, one has to fix
$$
\theta_+ = \theta_- = 0, \eqno(34)
$$
and there remains 4 self-adjoint extension parameters: $\varphi_+$, $\tilde{\varphi}_+$, $\varphi_-$ and $\tilde{\varphi}_-$. 
Operator $H$ (4) acting on functions which are defined in the region bounded by two parallel planes is self-adjoint, if the 
following condition holds:
$$
\left\{I - \beta\exp[{\rm i}(\varphi_{\pm}\gamma^{5} \pm
\tilde{\varphi}_{\pm}\alpha^{z})]\right\}\chi|_{z=\pm{a/2}}=0
\eqno(35)
$$
(the same condition holds for $\tilde{\chi}$). The latter ensures
the fulfilment of constraints
$$
\tilde{\chi}^{\dag}\alpha^{z}\chi|_{z=\pm{a/2}}=0 \eqno(36)
$$
and
$$
\tilde{\chi}^{\dag}\beta\exp\left\{{\rm i}[\varphi_{\pm}\gamma^{5}
\pm (\tilde{\varphi}_{\pm}
+\pi/2)\alpha^{z}]\right\}\chi|_{z=\pm{a/2}}=0.\eqno (37)
$$

\section{Induced vacuum energy in the magnetic field background}

The operator of a spinor field which is quantized in an ultrastatic
background is presented in the form
$$\hat{\Psi}(t,\mathbf{r})=\sum\!\!\!\!\!\!\!\!\!\!\!\int\limits_{E_{\lambda}>0}{\rm e}^{-{\rm i}E_{\lambda}t}\psi_{\lambda}(\mathbf{r})\hat{a}_{\lambda}
+\sum\!\!\!\!\!\!\!\!\!\!\!\int\limits_{E_{\lambda}<0}{\rm
e}^{-{\rm
i}E_{\lambda}t}\psi_{\lambda}(\mathbf{r})\hat{b}^{\dag}_{\lambda},\eqno(38)
$$
where
$\hat{a}^{\dag}_{\lambda}$ and $\hat{a}_{\lambda}$
($\hat{b}^{\dag}_{\lambda}$ and $\hat{b}_{\lambda}$) are the
spinor particle (antiparticle) creation and destruction operators,
satisfying anticommutation relations
$[\hat{a}_\lambda,\hat{a}_{\lambda'}^\dagger]_+=[\hat{b}_\lambda,\hat{b}_{\lambda'}^\dagger]_+=\left\langle
\lambda|\lambda'\right\rangle$,
wave functions $\psi_{\lambda}(\textbf{r})$ form a
complete set of solutions to the stationary Dirac equation
$$
H\psi_{\lambda}(\mathbf{r})=E_{\lambda}\psi_{\lambda}(\mathbf{r});\eqno(39)
$$
$\lambda$ is the set of parameters (quantum numbers) specifying a
one-particle state with energy $E_{\lambda}$; symbol
$\sum\!\!\!\!\!\!\!\int\,$ denotes summation over discrete and
integration (with a certain measure) over continuous values of
$\lambda$. Ground state $|\texttt{vac}>$ is defined by condition
$\hat{a}_\lambda|\texttt{vac}>=\hat{b}_\lambda|\texttt{vac}>=0$.
The temporal component of the operator of the energy-momentum tensor
is given by expression
$$
\hat{T}^{00}=\frac{\rm{i}}{4}[\hat{\Psi}^{\dag}({\partial_0}\hat{\Psi})-({\partial_0}\hat{\Psi}^{T})\hat{\Psi}^{{\dag}T}
-({\partial_0}\hat{\Psi}^{\dag})\hat{\Psi}+\hat{\Psi}^{T}({\partial_0}\hat{\Psi}^{{\dag}T})],\eqno(40)
$$
where superscript $T$ denotes a transposed spinor. Consequently, the formal expression for the vacuum expectation value of the energy
density is
$$
\varepsilon=<\texttt{vac}|\hat{T}^{00}|\texttt{vac}>=
-\frac{1}{2}\sum\!\!\!\!\!\!\!\!\!\int \,|E_{\lambda}|\psi_{\lambda}^{\dag}(\textbf{r})\psi_{\lambda}(\textbf{r}).\eqno(41)
$$

Let us consider the quantized charged massive spinor field
in the background of a static uniform magnetic field; then $\bnabla=\bpartial-{\rm
i}e\mathbf{A}$
and the connection can be chosen as $\mathbf{A}=(-yB,0,0)$,
where $B$ is the value of the magnetic field strength which is directed along the
$z$-axis in Cartesian coordinates $\mathbf{r}=(x,y,z)$, $\mathbf{B}=(0,0,B)$. The
one-particle energy spectrum is
$$
E_{nk}=\pm\omega_{nk},\eqno(42)
$$
where
$$
\omega_{nk}=\sqrt{2n|eB|+k^{2}+m^{2}},\;-\infty<k<\infty,\;n=0,1,2,...\,
,\eqno(43)
$$
$k$ is the value of the wave number vector along the $z$-axis, and
$n$ labels the Landau levels. Using the explicit form of the complete set of solutions to the Dirac equation, 
one can get that expression (41) takes form
$$
\varepsilon^{\infty}=-\frac{|eB|}{2\pi^{2}}\int\limits_{-\infty}^{\infty}{\rm d}k
\sum\limits_{n=0}^{\infty}\iota_{n}\omega_{nk},\eqno(44)
$$
where the superscript on the left-hand side indicates that the magnetic field fills the whole (infinite) space; 
the appearance of factor $\iota_{n}=1-\frac{1}{2} \delta_{n0}$ on the right-hand side is due to the fact that there is one 
solution for the lowest Landau level, $\psi^{(0)}_{q0k}(\mathbf{r})$ ($q$ is the value of the wave number vector along 
the $x$-axis, $-\infty<q<\infty$), and there are two solutions otherwise, 
$\psi^{(j)}_{qnk}(\mathbf{r})$ ($j=1,2$), $n\geq1$. 
The integral and the sum in (44) are divergent and
require regularization and renormalization. This problem has been
solved long ago by Heisenberg and Euler \cite{Hei2} (see also
\cite{Schw}), and we just list here their result
$$
\varepsilon^{\infty}_{\rm ren}=\frac{1}{8\pi^{2}}\int\limits_{0}^{\infty}\frac{{\rm d}\tau}{\tau}
{\rm e}^{-\tau}\left[\frac{eBm^{2}}{\tau}\coth\left(\frac{eB\tau}{m^{2}}\right)-\frac{m^{4}}{\tau^2}
-\frac{1}{3}e^{2}B^{2}\right];\eqno(45)
$$
note that the renormalization procedure involves subtraction at
$B=0$ and renormalization of the charge.

Let us turn now to the quantized charged massive spinor field in
the background of a static uniform magnetic field in spatial
region $\Omega$ bounded by two parallel planes
$\partial{\Omega}^{(+)}$ and $\partial{\Omega}^{(-)}$; the
position of $\partial{\Omega}^{(\pm)}$ is identified with
$z=\pm{a/2}$, and the magnetic field is orthogonal to the
boundary. The solution to (39) in region $\Omega$ is chosen as a
superposition of two plane waves propagating in opposite directions along the $z$-axis,
$$
\psi^{(j)}_{qnl}(\mathbf{r})=\psi^{(j)}_{qnk_{l}}(\mathbf{r})+\psi^{(j)}_{qn-k_{l}}(\mathbf{r}), \quad j=0,1,2, \eqno(46)
$$
where the values of wave number vector $k_{l} \; (l=0,\pm1,\pm2, ... )$ are
determined from the boundary condition, see (35):
$$
\left\{I-\beta\exp[{\rm i}(\varphi_{\pm}\gamma^{5} \pm
\tilde{\varphi}_{\pm}\alpha^{z})]\right\}\psi^{(j)}_{qnl}(\mathbf{r})|_{z=\pm{a/2}}=0,
\quad (j=1,2)  \quad  n\geq1 \eqno(47)
$$
and
$$
\biggl[I+\frac{\beta}{2}\biggl(\pm\alpha^{z}\gamma^{5}-1\biggr){\rm e}^{{\rm
i}(\varphi_{\pm}-\tilde\varphi_{\pm})\gamma^{5}}\Theta(\pm eB)
-\frac{\beta}{2}\biggl(\pm\alpha^{z}\gamma^{5}+1\biggr){\rm e}^{{\rm
i}(\varphi_{\pm}+\tilde\varphi_{\pm})\gamma^{5}}\Theta(\mp
eB)\biggr]\psi^{(0)}_{q0l}(\mathbf{r})|_{z=\pm{a/2}} =0; \eqno(48)
$$
the step function is introduced as $\Theta(u)=1$ at $u>0$
and $\Theta(u)=0$ at $u<0$. This boundary condition ensures that the normal component of current
$\mathbf{J}_{qnlj}(\mathbf{r})=\psi^{(j)\dag}_{qnl}(\mathbf{r})\balpha\psi^{(j)}_{qnl}(\mathbf{r})$ 
$(j=0,1,2)$ vanishes at the boundary, see (36),
$$
J^{z}_{qnlj}(\mathbf{r})|_{z=\pm{a/2}}=0, \eqno(49)
$$
which, cf. (1), signifies that the quantized matter is confined within the
boundaries.

The boundary condition depends on four self-adjoint extension parameters,
$\varphi_{+},\tilde{\varphi}_{+},\varphi_{-}$ and
$\tilde{\varphi}_{-}$, in the case of $n\geq1$, see (47), and on
two self-adjoint extension parameters, $\varphi_{+}-\tilde{\varphi}_{+}$ and
$\varphi_{-}+\tilde{\varphi}_{-}$  ($eB>0$), or $\varphi_{+}+\tilde{\varphi}_{+}$ and
$\varphi_{-}-\tilde{\varphi}_{-}$  ($eB<0$), in the case of $n=0$, see (48). As was mentioned in the previous section, 
the values of these self-adjoint extension parameters may
vary arbitrarily from point to point of the bounding planes.
However, in the context of the Casimir effect, such a generality seems to be excessive, lacking physical
motivation and, moreover, being impermissible, as long as boundary condition (47)-(48) is to be regarded as the 
one determining the spectrum of the wave number vector in the $z$-direction.
Therefore, we shall assume in the following that the self-adjoint
extension parameters are independent of coordinates $x$ and $y$.

It should be noted that value $k_l=0$ is allowed for special cases only. Really,
we have in the case of $k_l=0$:
$$
\psi_{qnl}^{(j)}({\bf r})|_{z=a/2}=\psi_{qnl}^{(j)}({\bf
r})|_{z=-a/2}, \eqno(50)
$$
and boundary condition (47)-(48) can be presented in the form
$$
R \, \psi_{qnl}^{(j)}({\bf r})|_{k_l=0}=0, \eqno(51)
$$
where
$$ \left\{
\begin{array}{l}R_{11}=\sin\frac{\varphi_+-\tilde{\varphi}_+}{2}, \quad
R_{12}=0, \quad R_{13}={\rm i}\cos\frac{\varphi_+-\tilde{\varphi}_+}{2}, \quad R_{14}=0, \\
R_{21}=0, \quad R_{22}=\sin\frac{\varphi_++\tilde{\varphi}_+}{2}, \quad R_{23}=0, \quad 
R_{24}={\rm i}\cos\frac{\varphi_++\tilde{\varphi}_+}{2}, \\
R_{31}=\sin\frac{\varphi_-+\tilde{\varphi}_-}{2}, \quad R_{32}=0,
\quad R_{33}={\rm i}\cos\frac{\varphi_-+\tilde{\varphi}_-}{2}, \quad R_{34}=0,
\\
R_{41}=0, \quad R_{42}=\sin\frac{\varphi_--\tilde{\varphi}_-}{2}, \quad R_{43}=0,
\quad R_{44}={\rm i}\cos\frac{\varphi_--\tilde{\varphi}_-}{2}
\end{array}\right\}. \eqno(52)
$$
The determinant of matrix $R$ is:
$$
\det R = -\sin\frac{\varphi_+-\varphi_-+\tilde{\varphi}_++\tilde{\varphi}_-}{2}
\sin\frac{\varphi_+-\varphi_--\tilde{\varphi}_+-\tilde{\varphi}_-}{2}. \eqno(53)
$$
The necessary and sufficient condition for value $k_l=0$ to be admissible is $\det R = 0$. 
Otherwise, $\det R \neq 0$ and value $k_l=0$ is excluded from the spectrum, 
because equation (51) then allows for the trivial solution only, $\psi_{qnl}^{(j)}({\bf r})|_{k_l=0} \equiv 0$.

The spectrum of $k_{l}$ is determined from a transcendental equation which in general possesses two branches and 
allows for complex values of $k_l$ (details will be published elsewhere). It is not clear which of the branches should be chosen, 
and, therefore, we restrict ourselves to boundary conditions corresponding to the case of a single branch. The latter is 
ensured by imposing constraint
$$
\varphi_{+}=\varphi_{-}=\varphi, \quad
\tilde{\varphi}_{+}=\tilde{\varphi}_{-}=\tilde{\varphi}. \eqno(54)
$$
Then relations (35) and (37) take forms
$$
\left\{I - \beta\exp[{\rm i}(\varphi\gamma^{5} \pm
\tilde{\varphi}\alpha^{z})]\right\}\chi|_{z=\pm{a/2}}=0
\eqno(55)
$$
and
$$
\tilde{\chi}^{\dag}\beta\exp\left\{{\rm i}[\varphi\gamma^{5}
\pm (\tilde{\varphi}
+\pi/2)\alpha^{z}]\right\}\chi|_{z=\pm{a/2}}=0.\eqno (56)
$$
respectively, while the equation determining the spectrum of $k_{l}$ takes form
$$
\cos(k_{l}a)+\frac{\omega_{nk_{l}}\,{\rm sgn}(E_{nk_{l}})\cos\tilde{\varphi}-m\cos\varphi}{k_{l}\sin\tilde{\varphi}}\sin(k_{l}a)
=0, \eqno(57)
$$
where ${\rm sgn}(u)=\Theta(u)-\Theta(-u)$ is the sign function; note that the spectrum is real, consisting of values of the 
same sign, say, $k_l>0$ (values of the opposite sign ($k_l<0$) should be excluded to avoid double counting).

In the case of $\tilde{\varphi}=- \pi/2$, the spectrum of $k_{l}$ is independent of the number of the Landau level, $n$, 
and of the sign of the one-particle energy, ${\rm sgn}(E_{nk_{l}})$; it is determined from equation  
$$
\cos(k_{l}a)+\frac{m\cos\varphi}{k_{l}}\sin(k_{l}a)=0. \eqno(58)
$$
In the case of $\tilde{\varphi}=0$, the $k_{l}$-spectrum is also independent of $n$ and of ${\rm sgn}(E_{nk_{l}})$; moreover, 
it is independent of $\varphi$, since the determining equation takes form
$$
\sin(k_{l}a)=0; \eqno(59)
$$
note that value $k_l=0$ is admissible in this case, see (53)-(54).
In what follows, we shall consider the most general case of two self-adjoint extension parameters, 
$\varphi$ and $\tilde{\varphi}$, when the $k_{l}$-spectrum depends on $n$ and on ${\rm sgn}(E_{nk_{l}})$, see (57).

Wave functions
$\psi^{(j)}_{qnl}(\mathbf{r})\,(j=0,1,2)$ satisfy the requirements
of orthonormality
$$\int\limits_{\Omega}{{\rm d}^{3}r}\,{\psi^{(j)\dag}_{qnl}(\mathbf{r})}\psi_{q'n'l'}^{(j')}(\mathbf{r})=
\delta_{jj'}\delta_{nn'}\delta_{ll'}\delta(q-q'), \quad j,j'=0,1,2\eqno(60)
$$
and completeness
$$
\sum\limits_{{\rm sgn}(E_{nk_{l}})}\int\limits_{-\infty}^{\infty}{\rm{d}}q\sum\limits_{l}\left[\psi^{(0)}_{q0l}(\mathbf{r})\psi^{(0)\dag}_{q0l}(\mathbf{r'})
+\sum\limits_{n=1}^{\infty}\sum\limits_{j=1,2}\psi^{(j)}_{qnl}(\mathbf{r})\psi^{(j)\dag}_{qnl}(\mathbf{r'})\right]=
I\delta(\mathbf{r}-\mathbf{r'}). \eqno(61)
$$
Consequently, we obtain the following expression for the
vacuum expectation value of the energy per unit area of the boundary
surface
$$
\frac{E}{S}\equiv\int\limits_{-a/2}^{a/2}{\rm{d}}z\,\varepsilon=
-\frac{|eB|}{2\pi}\sum\limits_{{\rm sgn}(E_{nk_{l}})}\sum\limits_{l}\sum\limits_{n=0}^{\infty}\iota_{n}\omega_{nk_{l}}.\eqno(62)
$$

\section{Casimir energy and force}

Expression (62) can be regarded as purely formal, since it is ill-defined due to the divergence of infinite sums 
over $l$ and $n$. To tame the divergence, a factor containing a regularization
parameter should be inserted in (62). A summation over values $k_{l}\geq 0$, which are
determined by (57), can be performed with the use of the Abel-Plana formula and its generalizations. 
In the cases of $\tilde{\varphi}=0$ and of $\varphi=-\tilde\varphi=\pi/2$, the well-known versions of the Abel-Plana
formula (see, e.g., \cite{Bor}),
$$
\left.\sum\limits_{{\rm sgn}(E_{nk_{l}})}\sum\limits_{k_{l}\geq 0}f(k_{l}^{2})\right|_{\sin(k_{l}a)=0}=
\frac{a}{\pi}\int\limits_{-\infty}^{\infty}{\rm{d}}k{f(k^{2})}-\frac{2{\rm
i}a}{\pi}\int\limits_{0}^{\infty}{\rm{d}}\kappa \frac{f[(-{\rm
i}\kappa)^{2}]-f[({\rm i}\kappa)^{2}]}{{\rm e}^{2{\kappa}a}-1}+f(0) \eqno(63)
$$
and
$$
\left.\sum\limits_{{\rm sgn}(E_{nk_{l}})}\sum\limits_{k_{l}>0}f(k_{l}^{2})\right|_{\cos(k_{l}a)=0}=
\frac{a}{\pi}\int\limits_{-\infty}^{\infty}{\rm{d}}k{f(k^{2})}+\frac{2{\rm
i}a}{\pi}\int\limits_{0}^{\infty}{\rm{d}}\kappa \frac{f[(-{\rm
i}\kappa)^{2}]-f[({\rm i}\kappa)^{2}]}{{\rm e}^{2{\kappa}a}+1},
\eqno(64)
$$
are used, respectively. Otherwise, we use the following version of the Abel-Plana formula:
$$
\sum\limits_{{\rm sgn}(E_{nk_{l}})}\sum\limits_{k_{l}>0}f(k_{l}^{2})=\frac{a}{\pi}\int\limits_{-\infty}^{\infty}{\rm{d}}k{f(k^{2})}
+\frac{2{\rm
i}a}{\pi}\int\limits_{0}^{\infty}{\rm{d}}\kappa\Lambda(\kappa)\{f[(-{\rm
i}\kappa)^{2}]-f[({\rm i}\kappa)^{2}]\}
$$
$$
-f(0)-\frac{1}{\pi}
\int\limits_{-\infty}^{\infty}{\rm{d}}k f(k^{2})
\frac{m\cos\varphi\sin\tilde\varphi[k^{2}-\mu_n(\varphi,\tilde\varphi)]}
{[k^{2}+\mu_n(\varphi,\tilde\varphi)]^2 +4k^{2}m^{2}\cos^{2}\varphi\sin^{2}\tilde\varphi},\eqno(65)
$$
where
$$
\Lambda(\kappa)=\Biggl( -[\kappa^2\cos2\tilde\varphi-\mu_n(\varphi,\tilde\varphi)]{\rm e}^{2{\kappa}a}
+{\kappa}^{2}+2{\kappa}m\cos\varphi\sin\tilde\varphi-\mu_n(\varphi,\tilde\varphi)
$$
$$
+\frac{\sin\tilde\varphi}{a}\left\{-\kappa^2 m\cos\varphi(\cos2\tilde\varphi{\rm e}^{2{\kappa}a}-1)
+[(2{\kappa}\sin\tilde\varphi-m\cos\varphi){\rm e}^{2{\kappa}a}\right.
$$
$$
\left.+m\cos\varphi]\mu_n(\varphi,\tilde\varphi)
\right\}[{\kappa}^{2}-2{\kappa}m\cos\varphi\sin\tilde\varphi-\mu_n(\varphi,\tilde\varphi)]^{-1}\Biggr)
$$
$$
\times
\left\{[{\kappa}^{2}-2{\kappa}m\cos\varphi\sin\tilde\varphi-\mu_n(\varphi,\tilde\varphi)]
{\rm e}^{4{\kappa}a}\right.
$$
$$
\left. -2[\kappa^2\cos2\tilde\varphi-\mu_n(\varphi,\tilde\varphi)]{\rm e}^{2{\kappa}a}
+{\kappa}^{2}+2{\kappa}m\cos\varphi\sin\tilde\varphi-\mu_n(\varphi,\tilde\varphi)\right\}^{-1} \eqno(66)
$$
and
$$
\mu_n(\varphi,\tilde\varphi)=2n|eB|\cos^{2}\tilde\varphi
+m^{2}\sin(\varphi+\tilde\varphi)\sin(\varphi-\tilde\varphi). \eqno(67)
$$
In (63)-(65), $f(u^{2})$ as a function of complex variable $u$ is assumed to decrease
sufficiently fast at large distances from the origin of the complex
$u$-plane, and this decrease is due to the use of some kind of regularization for (62). However, the regularization 
in the second integral on the right-hand side of (63)-(65) can be removed; then
$$
{\rm i}\{f[(-{\rm i}\kappa)^{2}]-f[({\rm
i}\kappa)^{2}]\}=-\frac{|eB|}{\pi}\sum\limits_{n=0}^{\infty}\iota_{n}\sqrt{\kappa^{2}-\omega^{2}_{n0}}\eqno(68)
$$
with the range of $\kappa$ restricted to
$\kappa>\omega_{n0}$ for the corresponding terms; here, recalling (43),
$\omega_{n0}=\sqrt{2n|eB|+m^{2}}$.
As to the first integral on the right-hand side of (63)-(65), one immediately
recognizes that it is equal to $\varepsilon^{\infty}$ (44)
multiplied by $a$. Hence, if one ignores for a moment the last term
of (63), as well as the terms in the last line of (65), then the problem of regularization and removal
of the divergency in expression (62) is the same as that in the case
of no boundaries, when the magnetic field fills the whole space.
Defining the Casimir energy as the vacuum energy per unit area of the boundary surface,
which is renormalized in the same way as in the case of no
boundaries, we obtain
$$
\frac{E_{\rm ren}}{S}=a\varepsilon^{\infty}_{\rm ren}-
\frac{2|eB|}{\pi^{2}}a\sum\limits_{n=0}^{\infty}\iota_{n}\int\limits_{\omega_{n0}}^{\infty}{\rm
d}\kappa\Lambda(\kappa)\sqrt{\kappa^{2}-\omega^{2}_{n0}}
$$
$$
+\frac{|eB|}{2\pi}\sum\limits_{n=0}^{\infty}\iota_{n}\omega_{n0}+\frac{|eB|}{2\pi^{2}}\int\limits_{-\infty}^{\infty}{\rm
d}k
\sum\limits_{n=0}^{\infty}\iota_{n}\sqrt{k^{2}+\omega^{2}_{n0}}
\frac{m\cos\varphi\sin\tilde\varphi[k^{2}-\mu_n(\varphi,\tilde\varphi)]}
{[k^{2}+\mu_n(\varphi,\tilde\varphi)]^2 +4k^{2}m^{2}\cos^{2}\varphi\sin^{2}\tilde\varphi},\eqno(69)
$$
$\varepsilon^{\infty}_{\rm ren}$ is given by (45). The sums and the integral in the last line of (69)
(which are due to the terms in the last line of (65) and which can be interpreted as describing the proper
energies of the boundary planes containing the sources of the
magnetic field) are divergent, but this divergency is
of no concern for us, because it has no physical consequences.
Rather than the Casimir energy, a physically relevant quantity is
the Casimir force per unit area of the boundary surface, i.e. pressure, which is defined as
$$
F=-\frac{\partial}{\partial a}\frac{E_{\rm ren}}{S},\eqno(70)
$$
and which is free from divergencies. We obtain
$$
F=-\varepsilon^{\infty}_{\rm ren}-
\frac{2|eB|}{\pi^{2}}\sum\limits_{n=0}^{\infty}\iota_{n}\int\limits_{\omega_{n0}}^{\infty}{\rm
d}\kappa\Upsilon(\kappa)\sqrt{\kappa^{2}-\omega_{n0}^{2}},\eqno(71)
$$
where
$$
\Upsilon(\kappa)\equiv-\frac{\partial}{\partial a}a\Lambda(\kappa)
=\left[\upsilon_{1}(\kappa){\rm e}^{6{\kappa}a}+\upsilon_{2}(\kappa){\rm e}^{4{\kappa}a}+
\upsilon_{3}(\kappa){\rm e}^{2{\kappa}a}+\upsilon_{4}(\kappa)\right]
$$
$$
\times\left\{[{\kappa}^{2}-2{\kappa}m\cos\varphi\sin\tilde\varphi-\mu_n(\varphi,\tilde\varphi)]
{\rm e}^{4{\kappa}a}\right.
$$
$$
\left. -2[\kappa^2\cos2\tilde\varphi-\mu_n(\varphi,\tilde\varphi)]{\rm e}^{2{\kappa}a}
+{\kappa}^{2}+2{\kappa}m\cos\varphi\sin\tilde\varphi-\mu_n(\varphi,\tilde\varphi)\right\}^{-2} \eqno(72)
$$
and
$$
\upsilon_{1}(\kappa)=-(2\kappa a-1)
[{\kappa}^{2}-2{\kappa}m\cos\varphi\sin\tilde\varphi-\mu_n(\varphi,\tilde\varphi)]
[\kappa^2\cos2\tilde\varphi-\mu_n(\varphi,\tilde\varphi)]
$$
$$
-2[{\kappa}^{2}m\cos\varphi\cos2\tilde\varphi
-(2{\kappa}\sin\tilde\varphi-m\cos\varphi)\mu_n(\varphi,\tilde\varphi)]\kappa\sin\tilde\varphi, \eqno(73)
$$
$$
\upsilon_{2}(\kappa)=(4\kappa a-3)
\left\{[{\kappa}^{2}-\mu_n(\varphi,\tilde\varphi)]^{2}
-4{\kappa}^{2}m^{2}\cos^{2}\varphi\sin^{2}\tilde\varphi\right\}
$$
$$
+8\kappa^2[\kappa^2\cos^{2}\tilde\varphi-m^{2}\cos^{2}\varphi-\mu_n(\varphi,\tilde\varphi)]\sin^{2}\tilde\varphi
%$$
%$$
+4[\kappa^2+\mu_n(\varphi,\tilde\varphi)]{\kappa}m\cos\varphi\sin\tilde\varphi, \eqno(74)
$$
$$
\upsilon_{3}(\kappa)=
-(2\kappa a-3)[{\kappa}^{2}+2{\kappa}m\cos\varphi\sin\tilde\varphi-\mu_n(\varphi,\tilde\varphi)]
[\kappa^2\cos2\tilde\varphi-\mu_n(\varphi,\tilde\varphi)]
$$
$$
-2[{\kappa}^{2}m\cos\varphi\cos2\tilde\varphi
+(2{\kappa}\sin\tilde\varphi+m\cos\varphi)\mu_n(\varphi,\tilde\varphi)]\kappa\sin\tilde\varphi, \eqno(75)
$$
$$
\upsilon_{4}(\kappa)=
-[{\kappa}^{2}+2{\kappa}m\cos\varphi\sin\tilde\varphi-\mu_n(\varphi,\tilde\varphi)]^{2}. \eqno(76) 
$$

In the case of $\tilde\varphi={-\pi}/{2}$, relations (72)-(76) are simplified: 
$$
\left.\Upsilon(\kappa)\right|_{\tilde\varphi={-\pi}/{2}}
=\frac{\left[\left(2\kappa a-1\right)
\left(\kappa^{2}-m^{2}\cos^{2}\varphi\right)-2{\kappa}m\cos\varphi\right]{\rm
e}^{2{\kappa}a}-\left(\kappa-m\cos\varphi\right)^{2}}
{\left[\left(\kappa+m\cos\varphi\right){\rm
e}^{2{\kappa}a}+\kappa-m\cos\varphi\right]^{2}}. \eqno(77)
$$
This case was exhaustively studied earlier \cite{Si1} in a different parametrization 
corresponding to substitution $\cos\varphi \rightarrow 1/\cosh\vartheta$; we remind here 
that the case of the MIT bag boundary condition is obtainable at $\varphi=0$ or $\vartheta=0$, respectively.

In the case of $\tilde\varphi=0$, we obtain
$$
\left.\Upsilon(\kappa)\right|_{\tilde\varphi=0}
=-\frac{\left(2\kappa a-1\right)
{\rm e}^{2{\kappa}a}+1}
{\left({\rm e}^{2{\kappa}a}-1\right)^{2}}. \eqno(78)
$$
The spectrum of $k_{l}$ in this case is explicitly given by
$k_{l}=\frac{\pi}{a}l \,\, (l=0,1,2,...)$, and the Casimir pressure takes form
$$
\left.F\right|_{\tilde\varphi=0}=-\varepsilon^{\infty}_{\rm
ren}+
\frac{2|eB|}{\pi^{2}}\sum\limits_{n=0}^{\infty}\iota_{n}\int\limits_{\omega_{n0}}^{\infty}\frac{{\rm
d}\kappa}{{\rm
e}^{2{\kappa}a}-1}\frac{\kappa^{2}}{\sqrt{\kappa^{2}-\omega_{n0}^{2}}}, \eqno(79)
$$
or, in the alternative representation,
$$
\left.F\right|_{\tilde\varphi=0}=-\varepsilon^{\infty}_{\rm
ren}+
\frac{2|eB|}{\pi^{2}}\sum\limits_{n=0}^{\infty}\iota_{n}\omega_{n0}^{2}\sum\limits_{j=1}^{\infty}\left[K_{0}(2j\omega_{n0}a)+
\frac{1}{2j\omega_{n0}a}K_{1}(2j\omega_{n0}a)\right], \eqno(80)
$$
where $K_{\rho}(s)$ is the Macdonald function of order $\rho$.

It is instructive to consider also the case of $\varphi=\pi/2$, when relations (72)-(76) are reduced to
$$
\left.\Upsilon(\kappa)\right|_{\varphi=\pi/2}
=-\biggl\{\biggl[\left(2\kappa a-1\right)\bigg(1-
\frac{2\kappa^{2}\sin^{2}\tilde\varphi}{\kappa^{2}-\omega_{n0}^{2}\cos^{2}\tilde\varphi}\biggr)
-\frac{\kappa^{2}\omega_{n0}^{2}\sin^{2}2\tilde\varphi}{\left(\kappa^{2}
-\omega_{n0}^{2}\cos^{2}\tilde\varphi\right)^{2}}\biggr]{\rm e}^{6{\kappa}a}
$$
$$
-\biggl[4\kappa a-3+
\frac{2\kappa^{2}\left(\kappa^{2}-\omega_{n0}^{2}\right)\sin^{2}2\tilde\varphi}{\left(\kappa^{2}
-\omega_{n0}^{2}\cos^{2}\tilde\varphi\right)^{2}}\biggr]{\rm e}^{4{\kappa}a}
$$
$$
+\biggl[\left(2\kappa a-3\right)\bigg(1-
\frac{2\kappa^{2}\sin^{2}\tilde\varphi}{\kappa^{2}-\omega_{n0}^{2}\cos^{2}\tilde\varphi}\biggr)
+\frac{\kappa^{2}\omega_{n0}^{2}\sin^{2}2\tilde\varphi}{\left(\kappa^{2}
-\omega_{n0}^{2}\cos^{2}\tilde\varphi\right)^{2}}\biggr]{\rm e}^{2{\kappa}a}+1\biggr\}
$$
$$
\times\biggl[\left({\rm e}^{2{\kappa}a}-1\right)^{2}+
\frac{4\kappa^{2}\sin^{2}\tilde\varphi}{\kappa^{2}-\omega_{n0}^{2}\cos^{2}\tilde\varphi}{\rm e}^{2{\kappa}a}\biggr]^{-2}. \eqno(81)
$$
This case interpolates between the case of spectrum $k_{l}=\frac{\pi}{a}l \,\, (l=0,1,2,...)$, 
see (78)-(80), and the case of 
spectrum $k_{l}=\frac{\pi}{a}(l+\frac{1}{2})$ $(l=0,1,2,...)$ with
$$
\left.\Upsilon(\kappa)\right|_{\varphi=-\tilde\varphi=\pi/2}
=\frac{\left(2\kappa a-1\right)
{\rm e}^{2{\kappa}a}-1}
{\left({\rm e}^{2{\kappa}a}+1\right)^{2}} \eqno(82)
$$
and
$$
\left.F\right|_{\varphi=-\tilde\varphi=\pi/2}=-\varepsilon^{\infty}_{\rm
ren}-
\frac{2|eB|}{\pi^{2}}\sum\limits_{n=0}^{\infty}\iota_{n}\int\limits_{\omega_{n0}}^{\infty}\frac{{\rm
d}\kappa}{{\rm
e}^{2{\kappa}a}+1}\frac{\kappa^{2}}{\sqrt{\kappa^{2}-\omega_{n0}^{2}}}, \eqno(83)
$$
or, alternatively,
$$
\left.F\right|_{\varphi=-\tilde\varphi=\pi/2}=-\varepsilon^{\infty}_{\rm
ren}
-\frac{2|eB|}{\pi^{2}}\sum\limits_{n=0}^{\infty}\iota_{n}\omega_{n0}^{2}\sum\limits_{j=1}^{\infty}
(-1)^{j-1}\left[K_{0}(2j\omega_{n0}a)+
\frac{1}{2j\omega_{n0}a}K_{1}(2j\omega_{n0}a)\right]. \eqno(84)
$$

\section{Asymptotics at small and large separations of plates}

The expression for the Casimir pressure, see (71), can be presented as 
$$
F=-\varepsilon^{\infty}_{\rm ren} + \Delta_{\varphi,\tilde\varphi}(a),  \eqno(85)
$$
where the first term is equal to minus the vacuum energy density which is induced by the magnetic field in  
unbounded space, see (45), whereas the second term which is given by the sum over $n$ and the integral over $\kappa$ in (71) 
depends on the distance between bounding plates and on a choice of boundary conditions at the plates. 

In the case of a weak magnetic field, $|B|\ll{m^{2}|e|^{-1}}$, substituting the sum by integral 
$\int\limits_{0}^{\infty}{\rm d}n$ and changing the integration variable, we get
$$
\Delta_{\varphi,\tilde\varphi}(a)=-
\frac{1}{\pi^{2}}\int\limits_{m}^{\infty}{\rm
d}\kappa(\kappa^{2}-m^{2})^{3/2}\int\limits_{0}^{1}{\rm
d}v\sqrt{1-v}\tilde\Upsilon(\kappa,v),\quad
|eB|\ll{m^{2}},\eqno(86)
$$
where $\tilde\Upsilon(\kappa,v)$ is obtained from $\Upsilon(\kappa)$ (72) by substitution
$\mu_n(\varphi,\tilde\varphi)\rightarrow\tilde\mu_{v,\kappa^{2}}(\varphi,\tilde\varphi)$ with
$$
\tilde\mu_{v,\kappa^{2}}(\varphi,\tilde\varphi)=v(\kappa^{2}-m^{2})\cos^{2}\tilde\varphi
+m^{2}\sin(\varphi+\tilde\varphi)\sin(\varphi-\tilde\varphi). \eqno(87)
$$
In the limit of small distances between the plates, $ma\ll1$, (86) becomes independent of the $\varphi$-parameter:
$$
\Delta_{\varphi,\tilde\varphi}(a)=
\frac{1}{4a^{4}}\biggl\{\frac{\pi^{2}}{30} - \int\limits_{0}^{1}{\rm d}v \, \rho_{\tilde\varphi}(v)
\biggl(1 - \frac{|\rho_{\tilde\varphi}(v)|}{\pi}\biggr)\biggl[\frac{3}{2}\sqrt{1-v} \, \rho_{\tilde\varphi}(v)
\biggl(1 - \frac{|\rho_{\tilde\varphi}(v)|}{\pi}\biggr)
$$
$$
+ \frac{v\sin2\tilde\varphi}{1-v\cos^{2}\tilde\varphi}
\biggl(\frac{1}{2} - \frac{|\rho_{\tilde\varphi}(v)|}{\pi}\biggr)\biggr]\biggr\}, \quad \sqrt{|eB|}a \ll ma \ll 1, \eqno(88)
$$
where
$$
\rho_{\tilde\varphi}(v)= \arcsin \biggl(\frac{\sin\tilde\varphi}{\sqrt{1-v\cos^{2}\tilde\varphi}}\biggr). \eqno(89)
$$
Thus, $\Delta_{\varphi,\tilde\varphi}(a)$ in this case is power-dependent on the distance between the plates as $a^{-4}$ with 
the dimensionless constant of proportionality, either positive or negative, depending on the value of the 
$\tilde\varphi$-parameter. In particular, we get
$$
\Delta_{\varphi,0}(a)=\frac{\pi^{2}}{120}\frac{1}{a^{4}}, \quad \sqrt{|eB|}a \ll ma \ll 1 \eqno(90)
$$
and
$$
\Delta_{\varphi,{-\pi}/{2}}(a)=-\frac{7}{8}\frac{\pi^{2}}{120}\frac{1}{a^{4}}, 
\quad \sqrt{|eB|}a \ll ma \ll 1. \eqno(91)
$$
In the limit of large distances between the plates, $ma\gg1$, $\Delta_{\varphi,\tilde\varphi}(a)$ (86) takes form
$$
\Delta_{\varphi,\tilde\varphi}(a)=
\frac{2}{\pi^{2}}\int\limits_{m}^{\infty}{\rm
d}\kappa\kappa(\kappa^{2}-m^{2})^{3/2}{\rm e}^{-2{\kappa}a}\int\limits_{0}^{1}{\rm
d}v\sqrt{1-v}
\biggl\{a\frac{\kappa^2\cos2\tilde\varphi-\tilde\mu_{v,\kappa^2}(\varphi,\tilde\varphi)}
{{\kappa}^{2}-2{\kappa}m\cos\varphi\sin\tilde\varphi-\tilde\mu_{v,\kappa^2}(\varphi,\tilde\varphi)}
$$
$$
-\frac{(2{\kappa}\sin\tilde\varphi-m\cos\varphi)\tilde\mu_{v,\kappa^2}(\varphi,\tilde\varphi)
-{\kappa}^{2}m\cos\varphi\cos2\tilde\varphi}
{[{\kappa}^{2}-2{\kappa}m\cos\varphi\sin\tilde\varphi-\tilde\mu_{v,\kappa^2}(\varphi,\tilde\varphi)]^{2}}\sin\tilde\varphi\biggr\},
 \quad |eB|\ll{m^{2}}, \quad  ma\gg1. \eqno(92)
$$
Clearly, (92) is suppressed as ${\rm exp}(-2ma)$. In particular, we get
$$
\Delta_{\varphi,0}(a)=\frac{1}{2\pi^{3/2}}\frac{m^{5/2}}{a^{3/2}}{\rm
e}^{-2ma}\left[1+O\left(\frac{1}{ma}\right)\right], \quad |eB|\ll{m^{2}}, \quad  ma\gg1  \eqno(93)
$$
and
$$
\Delta_{\varphi,{-\pi}/{2}}(a)=
\left\{\begin{array}{l}-\frac{3}{16\pi^{3/2}}\frac{m^{3/2}}{a^{5/2}}{\rm
e}^{-2ma}[1+O(\frac{1}{ma})],\quad \varphi=0
\\ [6 mm]
-\frac{\tan^{2}(\varphi/2)}{2\pi^{3/2}}\frac{m^{5/2}}{a^{3/2}}{\rm
e}^{-2ma}[1+O(\frac{1}{ma})],\quad \varphi \neq 0
\end{array}
\right\}, 
 \quad |eB|\ll{m^{2}}, \quad  ma\gg1.  \eqno(94)
$$

In the case of a strong magnetic field, $|B|\gg{m}^{2}|e|^{-1}$, one has
$$
\Delta_{\varphi,\tilde\varphi}(a)=-
\frac{|eB|}{\pi^{2}}\left[\int\limits_{m}^{\infty}{\rm
d}\kappa\sqrt{\kappa^{2}-m^{2}}\Upsilon(\kappa)|_{n=0}\right.
$$
$$
\left. +2\sum\limits_{n=1}^{\infty}\int\limits_{\sqrt{2n|eB|}}^{\infty}{\rm d}\kappa\sqrt{\kappa^{2}-2n|eB|}
\Upsilon(\kappa)|_{m=0}\right], \quad |eB|\gg{m}^{2}. \eqno(95)
$$
In the limit of extremely small distances between the plates, 
$ma\ll\sqrt{|eB|}a\ll1$, the analysis is 
similar to that of the limit of 
$\sqrt{|eB|}a \ll ma \ll 1$, yielding the same results as (88)-(91). Otherwise, in the 
limit of $\sqrt{|eB|}a\gg1$, only the first term in square brackets on the right-hand side of (95) matters. In the limit 
of small distances between the plates this term becomes $\varphi$-independent, yielding
$$
\Delta_{\varphi,\tilde\varphi}(a)=
\frac{|eB|}{4a^{2}}\left[\frac{1}{6}-\frac{|\tilde\varphi|}{\pi}\left(1-\frac{|\tilde\varphi|}{\pi}\right)\right], 
\quad \sqrt{|eB|}a\gg1,  \quad  ma\ll1. \eqno(96)
$$
In particular, we get
$$
\Delta_{\varphi,0}(a)=\frac{|eB|}{24a^{2}}, \quad
\sqrt{|eB|}a \gg 1,  \quad  ma \ll 1, \eqno(97)
$$
$$
\Delta_{\varphi,{\pm\pi}/{4}}(a)=- \frac{|eB|}{192a^{2}},\quad
\sqrt{|eB|}a\gg1,  \quad  ma\ll1 \eqno(98)
$$
and
$$
\Delta_{\varphi,{-\pi}/{2}}(a)=- \frac{|eB|}{48a^{2}},\quad
\sqrt{|eB|}a\gg1,  \quad  ma\ll1. \eqno(99)
$$
In the limit of large distances between the plates, the first term in square brackets on the right-hand side of (95) yields
$$
\Delta_{\varphi,\tilde\varphi}(a)=
\frac{2|eB|}{\pi^{2}}\int\limits_{m}^{\infty}{\rm
d}\kappa\kappa(\kappa^{2}-m^{2})^{1/2}{\rm e}^{-2{\kappa}a}
\biggl\{a\frac{\kappa^2\cos2\tilde\varphi-m^{2}\sin(\varphi+\tilde\varphi)\sin(\varphi-\tilde\varphi)}
{{\kappa}^{2}-2{\kappa}m\cos\varphi\sin\tilde\varphi-m^{2}\sin(\varphi+\tilde\varphi)\sin(\varphi-\tilde\varphi)}
$$
$$
+\frac{{\kappa}^{2}m\cos\varphi\cos2\tilde\varphi
-(2{\kappa}\sin\tilde\varphi-m\cos\varphi)m^{2}\sin(\varphi+\tilde\varphi)\sin(\varphi-\tilde\varphi)}
{[{\kappa}^{2}-2{\kappa}m\cos\varphi\sin\tilde\varphi-m^{2}\sin(\varphi+\tilde\varphi)\sin(\varphi-\tilde\varphi)]^{2}}
\sin\tilde\varphi\biggr\},
 \quad \sqrt{|eB|}a\gg{ma}\gg1, \eqno(100)
$$
which is obviously suppressed as ${\rm exp}(-2ma)$. In particular, we get
$$
\Delta_{\varphi,0}(a)=
\frac{|eB|}{2\pi^{3/2}}\frac{m^{3/2}}{a^{1/2}}{\rm
e}^{-2ma}\left[1+O\left(\frac{1}{ma}\right)\right],\quad
\sqrt{|eB|}a\gg{ma}\gg1 \eqno(101)
$$
and
$$
\Delta_{\varphi,{-\pi}/{2}}(a)=
\left\{\begin{array}{l}-\frac{|eB|}{16\pi^{3/2}}\frac{m^{1/2}}{a^{3/2}}{\rm
e}^{-2ma}[1+O(\frac{1}{ma})],\quad \varphi=0
\\ [6 mm]
-\frac{|eB|\tan^{2}(\varphi/2)}{2\pi^{3/2}}\frac{m^{3/2}}{a^{1/2}}{\rm
e}^{-2ma}[1+O(\frac{1}{ma})],\quad \varphi \neq 0 \end{array}
\right\},
 \quad \sqrt{|eB|}a\gg{ma}\gg1. \eqno(102)
$$

It is appropriate in this section to consider also the limiting case of $m\rightarrow0$. In view of the asymptotical behaviour 
of the boundary-independent piece of the Casimir pressure,
$$
-\varepsilon^{\infty}_{\rm
ren}=\frac{e^{2}B^{2}}{24\pi^{2}}\ln\frac{2|eB|}{m^{2}}, \quad m^{2}\ll|eB| \eqno(103)
$$
and
$$
-\varepsilon^{\infty}_{\rm
{ren}}=\frac{1}{360\pi^{2}}\frac{e^{4}B^{4}}{m^{4}}, \quad m^{2}\gg|eB|, \eqno(104)
$$
namely asymptotics (88) is relevant for this case, and the pressure from the vacuum of a confined massless 
spinor matter field is given by expression 
$$
\left.F\right|_{m=0, \, B=0}=
\frac{1}{8a^{4}}\biggl\{\frac{\pi^{2}}{30} - \int\limits_{0}^{1}{\rm
d}v \, \rho_{\tilde\varphi}(v)\biggl(1 - \frac{|\rho_{\tilde\varphi}(v)|}{\pi}\biggr)
$$
$$
\times\biggl[\frac{3}{2}\sqrt{1-v} \, \rho_{\tilde\varphi}(v)\biggl(1 - \frac{|\rho_{\tilde\varphi}(v)|}{\pi}\biggr) 
+ \frac{v\sin2\tilde\varphi}{1-v\cos^{2}\tilde\varphi}
\biggl(\frac{1}{2} - \frac{|\rho_{\tilde\varphi}(v)|}{\pi}\biggr)\biggr]\biggr\}, \eqno(105)
$$
which is bounded from above and below by values
$$
\left.F\right|_{m=0, \, B=0, \, \tilde\varphi= 0}=\frac{\pi^{2}}{240}\frac{1}{a^{4}} \eqno(106)
$$
and
$$
\left.F\right|_{m=0, \, B=0, \, \tilde\varphi={-\pi}/{2}}=-\frac{7}{8}\frac{\pi^{2}}{240}\frac{1}{a^{4}}, \eqno(107)
$$
respectively; here, an additional factor of 1/2 has appeared due to diminishment of the number of degrees 
of freedom (a massless spinor can be either left or right).

\section{Summary and discussion}

We study an influence of a background (classical) magnetic field on the
vacuum of a quantized charged spinor matter field which is confined
to a bounded region of space; the sources of the magnetic field are outside of the bounded region, and the magnetic 
field strength lines are assumed to be orthogonal to a boundary. The confinement of the matter field 
(i.e. absence of the matter flux across the boundary) is ensured by boundary condition (32) which is 
compatible with the self-adjointness of the Dirac hamiltonan operator and which generalizes the well-known 
MIT bag boundary condition to the most extent. In the case which is relevant to the geometry of 
the Casimir effect (i.e. the spatial region bounded by two parallel planes separated by distance $a$) and the uniform magnetic 
field orthogonal to the planes, the most general extension of the MIT bag boundary condition is given by (55); 
the spectrum of the wave number vector along 
the magnetic field in this case depends on the number of the Landau level and on the sign of the 
one-particle energy, see (57). The Casimir pressure is shown to take the form of (71), where  
$\varepsilon^{\infty}_{\rm {ren}}$ is given by (45) and $\Upsilon(\kappa)$ is given by (72)-(76); the result 
for the case of the MIT bag boundary condition is obtained from (71) at $\varphi=0$, $\tilde\varphi={-\pi}/{2}$. 

The Casimir effect is usually validated in experiments with nearly parallel plates separated by a distance of order 
$10^{-8}-10^{-5} \, \rm m$, see, e.g., \cite{Bor}. The pressure from the vacuum of neutral massless spinor matter 
onto the bounding plates is given by $a^{-4}$ times a constant of proportionality, either positive or negative, depending 
on a choice of boundary conditions, see (105)-(107). The situation with charged massive spinor matter is quite different. 
The Compton wavelength of the lightest charged particle, electron, is $m^{-1} \sim 10^{-12}\,\rm m$, thus ${ma}\gg1$ 
and, as has been shown in the preceding section, all the dependence of the Casimir pressure on the distance between the plates 
and a choice of boundary conditions at the plates is suppressed by factor ${\rm exp}(-2ma)$, see (92)-(94) and (100)-(102). 
Hence, the pressure from the electron-positron vacuum onto the plates separated by distance $a>10^{-10}\,\rm m$ 
is well approximated by $F \approx -\varepsilon^{\infty}_{\rm {ren}}$, where $\varepsilon^{\infty}_{\rm {ren}}$ (45) is 
negative, i.e. the pressure is positive and the plates are repelled. Some possibilities to detect this new-type Casimir 
effect were pointed out in \cite{Si1}.

\section*{Acknowledgments}

I would like to thank the Organizers of the XXIII International Conference on Integrable Systems and Quantum Symmetries for kind 
hospitality during this interesting and inspiring meeting. The work was supported by the National Academy of Sciences
of Ukraine (project No.0112U000054), the Program of Fundamental Research of the Department of Physics and
Astronomy of the National Academy of Sciences of Ukraine (project No.0112U000056) and the ICTP -- SEENET-MTP grant PRJ-09
``Strings and Cosmology''.

\end{document}